\documentstyle[12pt]{article}
\advance\textheight by 60pt
\advance\voffset by -40pt
\advance\textwidth by 50pt
\advance\oddsidemargin by -25pt
\advance\evensidemargin by -25pt

\def\single_space{\baselineskip 12pt plus 1pt minus 1pt}
\def\one_and_a_half_space{\baselineskip 19pt plus 1pt minus 1pt}
\def\double_spacesp{\baselineskip 25pt plus 2pt minus 2pt}

\def\atversim#1#2{\lower0.7ex\vbox{\baselineskip\zatskip
\lineskip\zatskip  \lineskiplimit 0pt\ialign{$\matth#1
\hfil##\hfil$\crcr#2\crcr\sim\crcr}}}

\begin{document}
\begin{titlepage}
\begin{flushright}
{\bf
PSU/TH/140\\
OCIP/C-94-1\\
January 1994\\
}
\end{flushright}
\vskip 1.5cm
{\Large
{\bf
\begin{center}
Aspects of $\chi$ and $\psi$ Production\\
in Polarized Proton-Proton Collisions \\
\end{center}
}}
\vskip 1.0cm
\begin{center}
M. A. Doncheski\\
Department of Physics\\
Carleton University\\
Ottawa, Ontario  K1S 5B6 Canada\\
\vskip .3cm
and\\
\vskip .3cm
R. W. Robinett \\
Department of Physics\\
The Pennsylvania State University\\
University Park, PA 16802 USA\\
\end{center}
\vskip 1.0cm
\begin{abstract}

Several topics of relevance to low transverse momentum $\psi$
and $\chi_{1,2}(c\overline{c})$ production in polarized
proton-proton collisions are discussed.  The leading
${\cal O}(\alpha_S^3)$ contributions to the low $p_T$
$\chi_1$ production
cross-sections via $gg$, $qg$, and $q\overline{q}$ initial
states are calculated as well as the corresponding spin-spin
asymmetries.  We find that $\chi_1$ production increases
relative to direct $\psi$ and $\chi_2$ production, providing
up to $25\%$ of the observable $e^+e^-$ pairs arising from
$\psi$ decays in $pp$
collisions at $\sqrt{s} = 500\,GeV$.  The spin-dependence of
$\chi_1$ production, however, is much smaller than for either
direct $\psi$ or $\chi_2$ production and so will likely be far
less useful than either process
in probing the polarized gluon structure function of
the proton.   A subset of the
${\cal O}(\alpha_S^3)$ radiative corrections to $\chi_2$ production
involving initial state quarks
are also performed and compared to leading order
$gg \rightarrow \chi_2$ predictions.

\end{abstract}
\end{titlepage}
\double_spacesp
\section{Introduction}

Even before the recent approval \cite{approval} of a program
of polarized proton-proton collisions at collider energies
at RHIC \cite{proposal,particleworld,workshop}, there had been a
rapid increase in the literature discussing the prospects for
probing the spin-dependence of many QCD processes and their
sensitivity to various polarized parton distributions in such
high energy polarized hadron collisions.

Familiar processes such as jet production \cite{jets1,jets2},
direct photon production \cite{jets1,qiu}, and weak boson
production \cite{soffernew,zphys} have all been examined in
this context in some detail.  For a few processes, such as
polarized Drell-Yan production \cite{ratcliffe} and direct
photon production \cite{conto2}, the full set of radiative
corrections have even been computed.  In these cases, the
next-to-leading (NLO) order corrections retain almost all of
the features of the leading-order (LO) calculations,
including their sensitivity to the sea quark and gluon
polarizations respectively.  On the other hand, heavy quark
production has recently been studied \cite{karliner} and it
seems likely that NLO corrections (especially at large $p_T$)
will change the spin-structure of $b$-quark production rather
dramatically, pointing out the importance of the comprehensive
study of NLO corrections to spin-dependent processes which might
be measured with the proposed RHIC detectors.

Quarkonium (especially $J/\psi$) production, both at low
transverse momentum \cite{cortes,Russians,conto1,doncheski} and at
high $p_T$ \cite{robinett,dkim},  has also been examined for
its possible spin-dependence and sensitivity to the polarized
gluon content of the proton.  One intriguing suggestion, due
to Cortes and Pire \cite{cortes} (hereafter CP), is to consider
low $p_T$ $\chi_2(c\overline{c})$ production where the dominant
lowest-order subprocess would be $gg \rightarrow \chi_2$.  To
this order, the observable polarization asymmetry,
\begin{equation}
A_{LL} = \frac{(\sigma(++)-\sigma(+-))}
              {(\sigma(++)+\sigma(+-))}\;\;\;\;\;,
\end{equation}
(where $\pm$ refers to the proton helicity) in such quantities
as $d\sigma/dy(y=0)$ (or $d\sigma/dx_F$) is directly proportional
to $(\Delta G(x)/G(x))^2$ times a partonic level spin-spin
asymmetry $\hat{a}_{LL}$.  Here $\Delta G(x) \equiv G_+(x) - G_-(x)$
is the longitudinally polarized gluon density, where $+/-$ refers
to a gluon with its helicity in the same/opposite direction as that
of the proton.  The partonic level asymmetries for $\chi_{2}/\chi_{0}$
production have been calculated in the context of potential models
\cite{doncheski} and are maximally large, $\hat{a}_{LL} = -1/+1$ for
$\chi_2/\chi_0$ respectively, as expected from simple angular
momentum conservation arguments. The study of the angular distribution
of the $\chi_2 \rightarrow \psi \gamma$ decay (which also provides the
clean signal) can simultaneously measure the value of $\hat{a}_{LL}$
\cite{cortes} as a further check. Similar experiments using $\chi_0$
production are hampered by the very small radiative branching ratio
to $\psi \gamma$ final states.  The spin-dependence of the other
dominant $\psi$ production process at low $p_T$, the so-called
`color-bleaching' mechanism, $g g \rightarrow g \psi$, has also been
examined \cite{doncheski}.

In the context of the RHIC spin physics program, the PHENIX detector
\cite{proposal}, with its excellent electron and photon detection,
will likely be able to resolve $\psi$ and $\chi_{1,2}$ states very
effectively so that further study of $\psi$ and $\chi_{1,2}$
production mechanisms at low $p_T$
in proton-proton collisions (polarized or not)
is quite timely.  (We note that the CDF Collaboration has presented
promising
preliminary results on such an analysis\cite{vaia}, demonstrating
the ability to separate (at some level) direct $\psi$ from radiative
$\chi_{1,2}$ decay in a high energy collider environment at large
transverse momentum.)

Questions can easily arise, however, concerning the reliability of
the theoretical understanding of $\chi(Q\overline{Q})$ production
mechanisms in hadronic collisions.  Issues such as the applicability
of potential model analyses of quarkonium production in hadronic
collisions at low values of $Q^2 = M_{\chi}^2$ (especially for
charmonium)  and,
as mentioned above, on the role of higher order QCD corrections are
obvious ones.  Both low transverse momentum inclusive $\psi$ and
$\Upsilon$ production \cite{martin,UA1} and high $p_T$ $\psi$
production \cite{halzen,GMS,UA12,mangano} have been well described by
such models which do, however, require relatively large $K$ factors,
namely $K \approx 2$, not unlike standard lowest-order jet, direct
photon and heavy quark analyses.  Recent calculations of radiative
corrections to low $p_T$ $^1S_0(Q\overline{Q})$ (i.e., $\eta_Q$)
production in hadron collisions \cite{kuhn}, albeit for toponium
production, confirm the presence of relatively large K-factors. These
corrections would presumably be even larger for $\eta_c$ states
because of the larger value of $\alpha_S$ at the lower charmonium
mass scale.
The
${\cal O}(\alpha_S^3)$ corrections to the inverse process, i.e.
$\chi_2 \rightarrow gg$ have been known for some time \cite{barbieri}
and were found to be quite large also indicating the possible
importance of higher-order radiative corrections in P-wave quarkonium
processes.  In addition, data on the relative yields of $\psi$ and
$\chi_{1,2}$ states in $pp$ collisions also seem to be compatible
with such potential models as well \cite{bauer}.

The presence of large multiplicative $K$-factors is, by itself, not a
strong argument against the use of the CP method as their analysis
relies simply on the $2 \rightarrow 1$ kinematic structure of the
$gg \rightarrow \chi_2$ subprocess.  Purely virtual corrections to
this process (retaining the $2 \rightarrow 1$ kinematics) are unlikely
to change the LO predictions for the spin-asymmetry which is dictated
by helicity conservation. The corrections arising from the
$2 \rightarrow 2$, $gg \rightarrow \chi_2 g$ process at the same order
(which are required to cancel the infrared divergences found in the
virtual diagrams)  can, however, change the kinematic structure of
$\chi_2$ production and hence the underlying spin-spin asymmetry.
Perhaps just as importantly, the NLO  ${\cal O}(\alpha_S^3)$ processes
$q g \rightarrow q \chi_2$ and $q \overline{q} \rightarrow g \chi_2$,
present at the same order, introduce a new dependence on the quark
and antiquark distributions so that even if the contribution of
quark/antiquark initiated processes is relatively small in the
{\it unpolarized} cross-section, it could make a significant impact
on the spin-dependence depending on the relative sizes of
$\Delta G(x)$ and $\Delta q(x)$.

A related process which appears for the first time at ${\cal O}(\alpha_S^3)$
is $\chi_1$ production.  Because the Yang-Landau-Pomeranchuk theorem
\cite{yang} forbids $gg \rightarrow \chi_1$ at lowest (or any) order,
the production of $\chi_1$ states begins at this order via the
processes $gg \rightarrow g \chi_1$, $qg \rightarrow q \chi_1$, and
$q \overline{q} \rightarrow q \chi_1$.  While suppressed relative to
$\chi_2$ production by an additional factor of $\alpha_S$, the
relatively large radiative branching ratio to observable $\psi$
states ($Br(\chi_1 \rightarrow \psi \gamma) = 0.27$ compared to
$Br(\chi_2 \rightarrow \psi \gamma) = 0.14$) partly compensates so we
expect $\chi_1$ production to be not unreasonably small.

Because direct $\psi$ production (via $gg \rightarrow g \psi$) and
$\chi_1$ production occur at the same order in $\alpha_S$, the ratio
of their production cross-sections may well be less sensitive to
assumptions about the momentum scale used in the perturbative
calculations and so could constitute a further test of a potential
model description of $\psi$ and $\chi$ state production as well as
allowing for an improved confrontation with data on
$\psi/\chi_1/\chi_2$ production ratios in hadroproduction experiments.
The variable center-of-mass energy available at RHIC ($\sqrt{s} =
50-500\,GeV$) will be quite useful in probing the energy dependence
of these ratios as well in probing the $x$-dependence of
$\Delta G(x)$.

In this work, we report on a set of NLO (i.e. ${\cal O}(\alpha_s^3)$)
calculations of relevance to these (and related) questions.  We
calculate a subset of the {\it spin-independent} radiative
corrections to $\chi_2$ production and include related results for
$\chi_0$ production although the small radiative decay branching
ratio to observable $\psi$ states make their production much less
useful.  Specifically we calculate the renormalized cross-sections
for the
subprocesses $qg \rightarrow q \chi_{0,2}$ and
$q \overline{q} \rightarrow g \chi_{0,2}$ to assess the relative
importance of initial states containing quarks and find that their
contribution is likely always smaller than $15\%$ of the total
cross-section.  This implies that the dominant processes for
$\chi_2$ production will still be from $gg$ initial states although
a full set of NLO corrections will likely be necessary for a confident
extraction of the polarized gluon distributions.  We stress that we
have {\it not} performed a complete set of radiative corrections to
$\chi_2$ production but we do feel that our partial results will be
useful in assessing the observability of various $\chi$ states and
the likely spin-dependence of their production for the RHIC spin
program.

Perhaps more importantly,
we also present all of the ${\cal O}(\alpha_S^3)$ contributions
to $\chi_1$ production, including their spin-dependence and find
that $\chi_1$ production increases in importance relative to direct
$\psi$ and $\chi_2$ production, providing up to $25\%$ of the lepton
pair yield from charmonium states at the highest RHIC energies.
The spin-dependence of $\chi_1$ production, however, is much
weaker than for either direct $\psi$ and $\chi_2$ production and
may not, therefore, contribute much to measurements of $\Delta G(x)$.

In the next section (2), we review the simplest $\psi$ and $\chi$
production mechanisms and describe the calculations of the partonic
level cross-sections for the $\chi_{1,2}$ processes we consider
while in Sec. 3 we present numerical results for $pp$ collisions
in the energy range accessible to RHIC (namely $50-500\,GeV$).
Finally, in Sec. 4, we discuss some aspects of the likely
spin-dependence of these new production mechanisms and make some
final comments.

\section{Partonic Level Cross-sections}

For reference, we begin by recalling the ${\cal O}(\alpha_S^2)$
$gg \rightarrow \chi_{0,2}$ cross-sections which form the basis for
the leading order description of $\chi$ production.  They are
well-known \cite{baier} and are given by
\begin{eqnarray}
\hat{\sigma}(gg \rightarrow \chi_0)
& = & \frac{12\pi^2\alpha_S^2|R_P'(0)|^2}{M^7}\delta(1-z)\\
\hat{\sigma}(gg \rightarrow \chi_2)
& = & \frac{16\pi^2\alpha_S^2|R_P'(0)|^2}{M^7}\delta(1-z)
\end{eqnarray}
where $z \equiv M^2/\hat{s}$ and $R_P'(0)$ is the derivative of
P-state radial wavefunction at the origin.  For completeness, we also
find it useful to quote the result for $gg \rightarrow g \psi$,
namely
\begin{equation}
\hat{\sigma}(gg \rightarrow g\psi) =
\frac{5 \pi \alpha_S^3 |R_S(0)|^2}{9 M^5} I(z)
\end{equation}
where
\begin{equation}
I(z) = 2z^2 \left[ \frac{1+z}{1-z} + \frac{2z \ln (z)}{(1-z)^2}
+ \frac{1-z}{(1+z)^2} - \frac{2z^2 \ln (z)}{(1+z)^3} \right]
\end{equation}
where $R_S(0)$ is the S-state radial wavefunction at the origin.  (We
note that, despite appearances, Eqn. 5 is finite as $z \rightarrow 1$.)

The first of the next-to-leading order corrections we consider are
those for $q \overline{q}  \rightarrow g \chi_J$ for $J = 0,1,2.$  The
matrix elements for these processes are readily available \cite{baier}
and can be integrated (over $d\hat{t}$) to yield the finite results
\begin{eqnarray}
\hat{\sigma}(q\overline{q} \rightarrow g \chi_0) & = &
\frac{128\pi\alpha_S^3|R_P'(0)|^2}{81M^7}
\frac{z^2(1-3z)^2}{(1-z)}\\
\hat{\sigma}(q\overline{q} \rightarrow g \chi_1) & = &
\frac{256\pi\alpha_S^3|R_P'(0)|^2}{27M^7}
\frac{z^2(1+z)}{(1-z)}\\
\hat{\sigma}(q\overline{q} \rightarrow g \chi_2) & = &
\frac{256\pi\alpha_S^3|R_P'(0)|^2}{81M^7}
\frac{z^2(1+3z+6z^2)}{(1-z)}.
\end{eqnarray}

The factors of $(1-z)^{-1}$ are not artifacts of any renormalization
but are symptomatic of P-wave bound state divergences which have been
noticed previously \cite{martin} and which can be regulated by
restricting the region of integration to $\hat{s} \geq (M + \Delta)^2$
where $\Delta$ is a typical binding energy, $\Delta \approx 0.3\, GeV$.
(Similar logarithmic divergences appear in the production of $D$-states
\cite{ernstrom} and in the multi-parton decays of P- and D-states
\cite{quigg,belanger} as well.)  We also note that the same kinematic
factors appear here as in the related processes
$Z^0 \rightarrow \chi_J + \gamma$ \cite{guberina}.  New techniques for
handling the related logarithms of binding energy which appear in
$P$-wave decays in a rigorous way
have been introduced \cite{bodwin} and could eventually
prove useful in this context as well.

The factor of $(1-3z)^2$ in Eqn.~6 implies a vanishing cross-section
at some physical center-of-mass energy which is not obviously related
to any symmetry.  (Unlike radiation zeros \cite{brown}, this zero
appears in the cross-section when integrated over angles.  As we will
see below, the zero may persist beyond tree level, also unlike the
case of radiation zeros.)  For future reference, we note that in each
of these cases, helicity conservation is sufficient to determine that
the partonic level asymmetries are $\hat{a}_{LL} = -1$.

The next subprocesses we consider are $q + g \rightarrow q + \chi_J$.
The cross-sections (in four dimensions) for these processes have also
been calculated \cite{baier} but when one attempts to integrate them
over angles one encounters infinities, at least in the case of
$\chi_0$ and $\chi_2$, which must be regulated.  The $\chi_1$ case is
more straightforward and we treat it first.  Because of Yang's theorem
\cite{yang}, when the $t$-channel gluon ($g^*$)
becomes soft the amplitude
describing $\chi_1 \leftrightarrow gg^*$ must vanish so that there is
no divergence in the phase space integral for this case.  The
(already) finite result is
\begin{equation}
\hat{\sigma}(qg\rightarrow q\chi_1) =
\frac{16\pi\alpha_S^3|R_P'(0)|^2}
             {9M^7} [(1-z)(5-4z-4z^2)-3z^2\ln(z)].
\end{equation}

For the $\chi_0$ and $\chi_2$ cases we must regularize the infinities
encountered in the angular integration and for this we use familiar
dimensional regularization techniques as in Ref. \cite{aem}. (A
complete set of calculations for the radiative corrections to
$gg \rightarrow \,^1S_0(Q\overline{Q})$ have also been carried out
using similar techniques \cite{kuhn}.)   The matrix elements for
$q g \rightarrow q \chi_{0,2}$, summed/averaged over all final/initial
polarizations, must then be calculated in $N = 4-2\epsilon$ dimensions.
After a long calculation using FORM \cite{form}, we find for the
$\chi_2$ case
\begin{equation}
\sum|{\cal M}|^2 = \frac{64\pi^2\alpha_S^3|R_P'(0)|^2}
                      {9M^3}
  \cdot \frac{F_4^{(2)}(s,t,u) + \epsilon\,F_{\epsilon}^{(2)}(s,t,u)}
           {(-t)(t-M^2)^4}
\end{equation}
where
\begin{equation}
F_4^{(2)}(s,t,u) = (t-M^2)^2(t^2+6M^4) - 2us(t^2-6M^2(t-M^2))
\end{equation}
and
\begin{equation}
F_{\epsilon}^{(2)}(s,t,u) = (t-M^2)[t^2(M^2-t) + 12(M^2(M^2-t)-us)]
\end{equation}
while for $\chi_0$ we have

\begin{equation}
\sum|{\cal M}|^2 = \frac{32\pi^2\alpha_S^3|R_P'(0)|^2}
                      {9M^3}
  \cdot    \frac{F_4^{(0)}(s,t,u) - \epsilon\,F_{\epsilon}^{(0)}(s,t,u)}
           {(-t)(t-M^2)^4}
\end{equation}
where
\begin{equation}
F_4^{(0)}(s,t,u) = (t-3M^2)^2(s^2+u^2)
\end{equation}
and
\begin{equation}
F_{\epsilon}^{(0)}(s,t,u) = (t-3M^2)^2(t-M^2)^2.
\end{equation}
To be consistent with more recent treatments of radiative corrections
we choose to average over the initial state spins by using
$2(1-\epsilon)$ degrees of freedom for the gluon which implies an
additional factor of $(1-\epsilon)^{-1}$ in Eqns. 10 and 13.  (This
choice has been discussed by Ellis and Sexton \cite{sexton} and is
used in Ref. \cite{kuhn}.)  Furthermore, we choose to work in the
$\overline{MS}$ scheme in which case the appropriate phase space
factor is \cite{ratcliffe}
\begin{equation}
PS = \frac{1}{8\pi}\left(\frac{1}{M^2}\right)^{\epsilon}
z^{\epsilon}(1-z)^{1-2\epsilon}
\int_{0}^{1}\,dy\,[y(1-y)]^{-\epsilon}
\end{equation}
where the subprocess invariants are
\begin{equation}
\hat{s} = \frac{M^2}{z}, \;\;\;\;\;\;\;
\hat{t} = -\frac{M^2}{z}(1-z)(1-y), \;\;\;
\hat{u} = -\frac{M^2}{z}(1-z)y.
\end{equation}

When one performs the angular integrals, the divergent $1/\epsilon$
term is proportional to the appropriate Altarelli-Parisi splitting
function ($P_{gq}(z)$ in this case) and is absorbed into the
scale-dependent parton distributions leaving a finite result.  Then
using the natural choice of scale $\mu = M$ we find the cross-sections
\begin{equation}
\hat{\sigma}(qg \rightarrow q\chi_2) = \frac{16\pi\alpha_S^3|R_P'(0)|^2}
                       {27M^7} G_2(z)
\end{equation}
where
\begin{eqnarray}
G_2(z) & = & 36(2-2z+z^2)\ln(1-z) - 3z^2\ln(z) \nonumber \\
     &   & -53 + 69z-18z^2+20z^3
\end{eqnarray}
and
\begin{equation}
\hat{\sigma}(qg \rightarrow q \chi_0) = \frac{8\pi\alpha_S^3|R_P'(0)|^2}
                       {27M^7} G_0(z)
\end{equation}
where
\begin{eqnarray}
G_0(z) & = & 54(2-2z+z^2)\ln(1-z) - 3z^2\ln(z) \nonumber \\
     &   & + 4(-35 + 42z - 9z^2 + 2z^3).
\end{eqnarray}

We first note that, after renormalization, these contributions to the
$\chi_2$ and $\chi_0$ cross-sections are, in fact, negative which is
quite similar to the case of Drell-Yan production \cite{aem} where the
$qg \rightarrow q\gamma^*$ correction to the tree-level process
$q\overline{q} \rightarrow \gamma^*$ is also negative.  The contribution
of the renormalized
$qg \rightarrow q \,^1S_0(Q\overline{Q})$ cross-section to
$\eta_Q$ production calculated in Ref. \cite{kuhn} is  similarly
negative.  We recall that the already finite $\chi_1$ contribution is
positive.

As an aside, we note that the $N = 4-2\epsilon$ matrix-element squared
for $q\overline{q} \rightarrow g\chi_0$ can be obtained from Eqns.
13--15 by crossing and one can see that the zero in the matrix element
at $z = 1/3$ (i.e. at $\hat{s} = 3M^2$) in $\chi_0$ production is
seemingly still present beyond tree level. We have no explanation for
this fact.

The final case we then consider are the diagrams leading to
$gg \rightarrow g\chi_J$.  For the case of $\chi_{0,2}$ these diagrams
must be combined with the virtual corrections to
$gg \rightarrow \chi_{0,2}$ to obtain a finite result and we do not
attempt a complete analysis of those diagrams here.  For the $\chi_1$
case, as there is no ${\cal O}(\alpha_S^2)$ contribution, the
$gg \rightarrow g\chi_1$ diagram gives the first non-vanishing
contribution and it is well-behaved when integrated over angles.  The
matrix element squared for $gg \rightarrow  g\chi_1$ has been calculated in
Ref. \cite{gastmans} and the total cross-section can be obtained by
directly integrating the four-dimensional results to find
\begin{equation}
\hat{\sigma}(gg \rightarrow g\chi_1) = \frac{4\pi\alpha_S^3|R_P'(0)|^2}
                  {M^7} \frac{[(1-z^2)H_1(z) + 12z^2\ln(z)H_1'(z)]}
                   {(1+z)^5(1-z)^4}
\end{equation}
where
\begin{eqnarray}
H_1(z) & = & z^9 + 39z^8 + 145z^7 + 251z^6 + 119z^5  \nonumber \\
       &   & - 153z^4 - 17z^3 - 147z^2 - 8z + 10
\end{eqnarray}
\begin{eqnarray}
H_1'(z) & = & z^8 + 9z^7 + 26z^6 + 28z^5 +17z^4\nonumber \\
       &   &  + 7z^3 - 40z^2 - 4z -4.
\end{eqnarray}
Because there is an $s$-channel contribution to this processes, one
obtains the bound-state divergences (the $1/(1-z)^4$ factor) which are
also present in the $q\overline{q} \rightarrow g\chi_J$ cases
considered earlier and which are handled in the same way.

\section{Numerical Results}

We can now use these cross-sections to evaluate the contributions of
quark (and antiquark) initiated processes compared to the tree-level
gluon-gluon fusion mechanism for $\chi_{0,2}$ states and all the
${\cal O}(\alpha_S^3)$ contributions for $\chi_1$ production.  In all our
calculations we use $\alpha_S(Q^2\!=\!M_{\chi}^2) = 0.26$
(corresponding to a leading-order $\Lambda = 200\, MeV$) and the
wavefunction values $|R_S(0)|^2 = 0.7 \, GeV^3$ and
$|R_P'(0)|^2/M_{\chi}^2 = 0.006\,GeV^3$ \cite{GMS}.  We use a recently
updated LO parameterization of parton distributions \cite{dukeowens}
evaluated at $Q^2 = M_{\chi}^2$ and the P-wave cutoff parameter
$\Delta = 0.3 \;GeV$.  The effects of changing $\Delta $ by
$\pm 0.1\,GeV$ is at most a few percent and so is not a major
source of theoretical uncertainty.  A dependence on $\Delta$ will
also arise in the renormalized $gg \rightarrow g \chi_{0,2}$
cross-sections and we expect a similar lack of sensitivity there
as well.

In Fig. 1a, we plot the contributions to $\chi_2$ production in $pp$
collisions corresponding to the LO $gg \rightarrow \chi_2$ and NLO
$qg \rightarrow q \chi_2$ and $q\overline{q} \rightarrow g \chi_2$
subprocesses.  Since the contribution from the renormalized $qg$
initial state is negative, its absolute value is plotted.  One can
see that the $q\overline{q}$ processes are nowhere important in the
RHIC energy range while the $qg$ initiated events comprise
approximately $15\%$ ($30\%$) of the total (Born) level $\chi_2$
cross-section (assuming a constant $K=2$ factor) which is not an
overwhelmingly large effect.  The contribution from $qg$ and
$q\overline{q}$ initial states is similar to that found for
$\eta_Q$ production found in Ref. \cite{kuhn} although the
contributions from $qg$ states is somewhat larger. This implies that
the total cross-section is still dominated by gluon fusion and so
can still provide information on the polarized gluon distribution.
A reliable extraction of $\Delta G(x)$, however, may well require,
however, a full set of spin-dependent NLO corrections.

The various (already finite and positive) contributions to $\chi_1$
production from Eqns. 7, 9, and 22, are evaluated and shown in Fig. 1b.
These results can then be combined with the direct $\psi$ production
mechanism to provide estimates of the yield of $e^+e^-$ pairs from
$\psi$, $\chi_2$, and $\chi_1$ states respectively once radiative
branching ratios for the $\chi_{1,2}$ states are included in Fig. 2.
  We see
that the relative amounts of $\psi$ and $\chi_2$ contributions to the
observable lepton pair cross-section stays roughly constant over the
RHIC range while the $\chi_1$ contributions increase to a rather
substantial fraction at the highest RHIC energies.  This fact offers
yet another motivation for the usefulness of the high resolution
electron and photon detection possible with the PHENIX detector at RHIC.
We recall that an empirical K-factor of roughly 2 is required for the
combined $\chi_2$ and $\psi$ contributions to the observble
$\psi$ production data so that the prediction for
$\chi_1$ contribution (as well as the other two)
might well be underestimated here by a factor of 2.

\section{Spin-spin Asymmetries}

One of the strongest motivations for considering quarkonium production
at RHIC is the possibility of measuring the spin-dependence of the
various production mechanisms, part of which includes their possible
sensitivity to the longitudinally polarized gluon distribution.  A full
treatment of the spin-dependent radiative corrections to $\chi_2$
production would be necessary to fully assess the extent to which the
suggestion of Cortes and Pire really provides a truly direct
measurement of $\Delta G$ but, barring that, some useful results can be
already be gleaned from our partial study.

Since for the $\chi_1$ case, all of the individual cross-sections were
finite at lowest order, we can make direct use of the individual
helicity amplitudes for the contributing processes to derive the
partonic-level spin-spin asymmetries for all of the ${\cal O}(\alpha_S^3)$
processes giving rise to $\chi_1$ production.

For the case of $ q g \rightarrow q \chi_1$, we can integrate the
individual helicity amplitudes found in Refs. \cite{robinett} and \cite{ubi} to
find
\begin{equation}
\hat{a}_{LL}(qg \rightarrow q \chi_1) = \frac{3z(1-z+z\ln(z))}
                 {(1-z)(5-4z-4z^2) - 3z^2\ln(z)}.
\end{equation}
The corresponding spin-spin asymmetry for the total
$g g \rightarrow g \chi_1$ cross-section is similarly calculable and is
given by
\begin{equation}
\hat{a}_{LL}(gg \rightarrow g\chi_1)
 = \frac{z[(1-z^2)A_1(z) + 12z\ln(z)A_1'(z)]}
                      {(z^2-1)H_1(z) - 12z^2\ln(z)H_1'(z)}
\end{equation}
where
\begin{eqnarray}
A_1(z) & = & z^8 + 33z^7 + 145z^6 + 271z^5 + 43z^4 \nonumber\\
       &   & +55z^3 - 273z^2 - 23z -12
\end{eqnarray}
\begin{eqnarray}
A_1'(z) & = & z^8 + 8z^7 + 25z^6 + 29z^5 + 34z^4 \nonumber \\
       &   & -32z^3 - 3z^2 - 21z -1.
\end{eqnarray}
and the $H_1(z)$ and $H_1'(z)$ are given in Eqns. 23 and 24.  As
mentioned above, all of the $q \overline{q} \rightarrow g \chi_{J}$
cross-sections have $\hat{a}_{LL} = -1$ from helicity conservation.
The first two of these asymmetries are plotted versus the natural
variable $z \equiv M^2/\hat{s}$ in Fig. 3 where it can be noted
that they
are positive, and clearly smaller in magnitude
than the maximally large value of
$\hat{a}_{LL} = -1$ for lowest order $\chi_2$ production.  For
comparison, we also plot the corresponding asymmetry for direct $\psi$
production, namely
\begin{equation}
\hat{a}_{LL} = -z \left[\frac{(1-z^2)F(z) + 2G(z) \ln(z)}
{(1-z^2)G(z) + 2z^2F(z) \ln (z)} \right]
\end{equation}
where
\begin{equation}
F(z) = z^2 + 2z + 5 \;\;\;\;\; \mbox{and} \;\;\;\;\;
G(z) = z^3 + 4z^2 + z + 2
\end{equation}
which is seen to be somewhat but not dramatically larger.

A useful measure of the average `analyzing power' or spin-dependence
in such a reaction is the average spin-spin asymmetry
\begin{equation}
<\hat{a}_{LL}> = \frac
{\sum_{ij}  \int dx_1 \int dx_2 f_i(x_1,Q^2) f_j(x_2,Q^2)
\hat{a}_{LL} d \hat{\sigma}}
{\sum_{ij}  \int dx_1 \int dx_2 f_i(x_1,Q^2) f_j(x_2,Q^2)
d \hat{\sigma}}
\end{equation}
which measures both the importance of the process to the total
cross-section and its spin-spin asymmetry. (If the unpolarized parton
densities, $f_{i,j}(x,Q^2)$, in the numerator were replaced by the
corresponding {\it polarized} distributions, $\Delta f_{i,j}(x,Q^2)$,
the result would be the observable spin-spin asymmetry.)  The average
spin-spin asymmetry for both $\chi_1$ production (including all three
contributing processes) as well as that for direct $\psi$ production
are shown in Fig. 4 where we see the spin-dependence of $\chi_1$
production is not at all large.  We comment that the average value
of $z \equiv M^2/\hat{s}$ for the $gg \rightarrow g \psi$
($gg \rightarrow g \chi_1$) process is roughly $0.5$ ($0.25-0.15$) over
the energy range plotted which accounts for the size and relatively
constant values of $<\!\hat{a}_{LL}\!>$.
So, while the measurement of $\chi_1$
production will provide another cross check on a potential model
description of charmonium production, it will not likely give much
information on the polarized gluon content of the proton.

As mentioned above, we have not evaluated the spin-dependent
renormalized cross-sections for $\chi_2$
production
(which in this case require a careful
treatment of the helicity amplitudes in arbitrary dimensions,
{\it i.e.} the $\gamma_5$ problem) but we can perhaps get some feel
for the partonic level asymmetries for $qg \rightarrow q\chi_{0,2}$
by examining the known $2 \rightarrow 2$ asymmetries \cite{robinett}
in the $\hat{t} \rightarrow 0$ limit.  While the partonic level
cross-sections are divergent in this limit, necessitating the
renormalization procedure above, the asymmetries are well-behaved. We
find in this limit (when $\hat{t} \rightarrow 0$ and
$\hat{u} \rightarrow (M^2-\hat{s})$)
\begin{equation}
\hat{a}_{LL}(qg \rightarrow q\chi_2) =
\frac{-2z+z^2}{2-2z+z^2} = -\hat{a}_{LL}(qg \rightarrow q\chi_0).
\end{equation}
where $z \equiv M^2/\hat{s}$.  We plot this function in Fig. 5 and note
that at threshold ($z=1$) the partonic level asymmetry is equal to
$-1 (+1)$ just as for $gg \rightarrow \chi_2 (\chi_0)$ at tree level.

For comparison, we also plot on the same figure the partonic level
asymmetry for $gg \rightarrow g\chi_2$, again in the same limit, and note
a similar behavior.  As before, the $\chi_{0,2}$ asymmetries are
trivially related and we find
\begin{equation}
\hat{a}_{LL}(gg \rightarrow g\chi_2) =
\frac{-z(2-3z+2z^2)}{(1-z+z^2)^2} =
-\hat{a}_{LL}(gg \rightarrow g\chi_0).
\end{equation}
The purely gluonic asymmetry, however, stays near its threshold value
of $-1$ over a wider range of $z$ and so will contribute more to the
observed asymmetry.  (These expressions are also found in the
$\hat{u} \rightarrow 0$ limit where divergences also occur in the
$gg \rightarrow g\chi_{0,2}$ processes.)  If the fully renormalized
gluon induced cross-section were described by these limiting values for
the asymmetries, then an average value of $<\!z\!>$ of
$0.5$ ($0.2$), as for $gg \rightarrow g  \psi$
($gg \rightarrow g \chi_1$),  would correspond
to $<\!\hat{a}_{LL}\!> \approx -0.9 \,(-0.5)$.
These results suggest that
the spin-dependence of the NLO $gg$ and $qg$ processes will be
somewhat `softer' but will still retain much of the
same general structure (most especially the sign) of the tree-level
spin-spin asymmetry in contrast to, for example, open heavy flavor
production as discussed in Ref. \cite{karliner}.

As a final comment, we note that $\chi_2(c\overline{c})$
production RHIC probes
the polarized gluon distributions in a kinematic region given by
\begin{displaymath}
\sqrt{\tau} \equiv \frac{M}{\sqrt{s}} \approx 0.07 - 0.007\;\;\;\;.
\end{displaymath}
We note that the production of a $^1S_0(t\overline{t})$ toponium
state (presumably detected via its two-photon decay) also has
a maximally large
partonic level spin-spin asymmetry given by $\hat{a}_{LL} = +1$.
A future supercollider (such as the LHC) with a polarization
option would be expected to probe  $\eta_t$ production in a
kinematic range given by $\sqrt{\tau} \approx .0017-.0025$ which would
then be in the region of polarized
gluon densities already probed by RHIC charmonium studies.
Since the value of $\alpha_S$ probed by such heavy mass states would
be much smaller than for charmonium, the effects of radiative corrections
would be expected to be much smaller and the interplay
between polarized gluon densities and the leading-order production
mechanisms would be expected to be much more direct.  In the same
context, standard model Higgs boson production via gluon fusion, i.e.
$gg \rightarrow H_0$ via quark loops, has the same leading-order
spin-spin asymmetry, namely $+1$. Higgs boson production at the LHC
via this mechanism would then be in the same kinematic region as
$\chi_2(c\overline{c})$ production at RHIC for Higgs boson  masses in the
range $200 - 2000 \, GeV$ which covers most of the expected range.  In
this regard, the spin-independent radiative correction calculations
of Refs. \cite{zerwas,dawson} are already useful in assessing the contributions
of various NLO processes.

\section{Acknowledgments}

We thank R. K. Ellis, J. Collins, A. Yokosawa, H. Grotch, G. Bodwin,
F. Halzen, and B. Kneihl for useful conversations.  This work was
supported in part by the National Science Foundation under grant
PHY--9001744 (R.R.), by the Texas National Research Laboratory
Commission under an SSC Junior Faculty Fellowship (R.R.) and by the
Natural Sciences and Engineering Research Council of Canada (NSERC)
(M.D.).

\newpage
{\Large
{\bf Figure Captions}}
\begin{itemize}
\item[Fig.\thinspace 1] The total cross-section, $\sigma(nb)$, for
$\chi_{2,1}(c\overline{c})$ production in pp collisions versus
$\sqrt{s}$ $(GeV)$. For 1(a), we plot the results for
$\chi_2$ production with the solid (dotted, dashed) curves corresponding
to the ${\cal O}(\alpha_S^2)$ $gg \rightarrow \chi_2$
($qg \rightarrow q \chi_2$, $q \overline{q} \rightarrow g \chi_2$)
contributions.  The {\it negative} of the $qg$ contribution is
plotted for simplicity. For 1(b), the results for $\chi_1$ production
are shown where the solid (dotted, dashed) curves correspond to
the $gg \rightarrow g \chi_1$ ($qg \rightarrow q \chi_1$,
$q \overline{q} \rightarrow g \chi_1$) contributions.

\item[Fig.\thinspace 2.] The total cross-section for $\psi$ production
(times leptonic branching ratio, $Br(\psi \rightarrow e^+e^-)$) in pp
collisions versus $\sqrt{s} (GeV)$.  The contributions from the lowest
order $gg \rightarrow g \psi$ (solid), ${\cal O}(\alpha_S^2)$
$gg \rightarrow \chi_2$ (dotted), and total ${\cal O}(\alpha_S^3)$ $\chi_1$
contributions (dashed) from $g$, $qg$, and $q\overline{q}$ initial
states are plotted.

\item[Fig.\thinspace 3.] The partonic level spin-spin asymmetry,
$\hat{a}_{LL}$, (in the total integrated cross-section) versus
$z \equiv M^2/\sqrt{s}$ for $\chi_1$ production.  The solid (dotted)
curves correspond to the asymmetry for $gg \rightarrow g \chi_1$
($qg \rightarrow q \chi_1$).  The asymmetry for direct $\psi$
production via $gg \rightarrow g \psi$ production (dashed curve) is
also shown for comparison.  The partonic level asymmetries for
$q \overline{q} \rightarrow g \chi_{0,1,2}$ and for
$gg \rightarrow \chi_2$ are both $\hat{a}_{LL} = -1$.

\item[Fig.\thinspace 4.] The average spin-spin asymmetry,
$<\!\hat{a}_{LL}\!>$ versus the center of mass energy,
$\sqrt{s}$ $GeV$ for $pp$ collisions.  The results for
total $\chi_1$ production (dots) (from $gg$, $qg$, and
$q\overline{q}$ processes) and direct $\psi$ production
(solid) (via $gg \rightarrow g \psi$) are shown.

\item[Fig.\thinspace 5.] The partonic level spin-spin asymmetries,
$\hat{a}_{LL}$,
versus $z \equiv M^2/\hat{s}$ for $\chi_2$ production in the limit
that $\hat{t} \rightarrow 0$.  The solid (dotted) curves correspond
to the processes $gg \rightarrow g \chi_2$
($qg \rightarrow q \chi_2$).

\end{itemize}

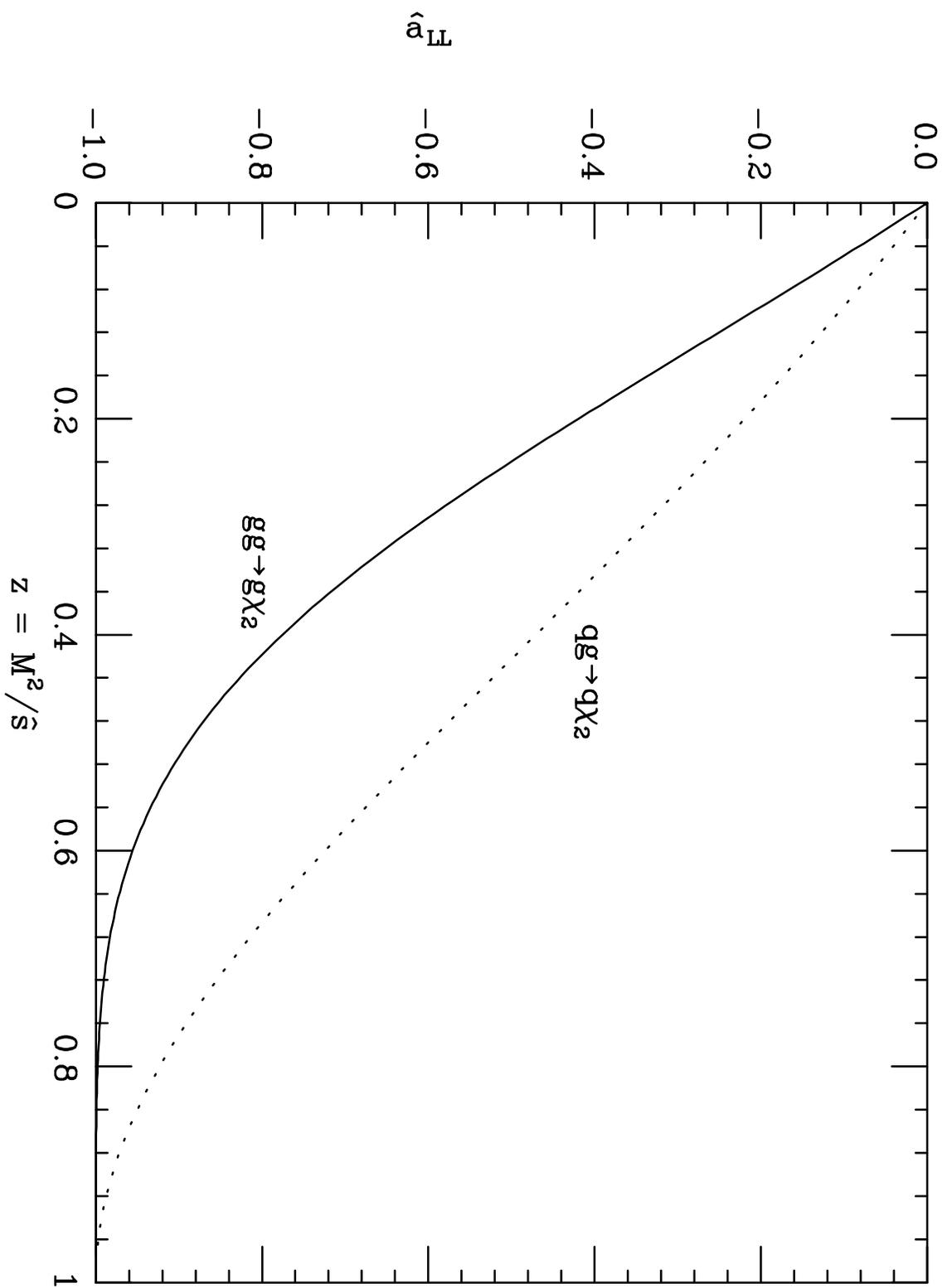
\end{document}